\documentclass{aa}

\usepackage{xcolor}
\usepackage{graphicx} 
\usepackage[per-mode=reciprocal, multi-part-units = single, range-units = single]{siunitx}		
\DeclareSIUnit[number-unit-product = {\,}]
  \Maxwell{Mx}
\DeclareSIUnit[number-unit-product = {\,}]
  \Gauss{Gauss}
\DeclareSIUnit[number-unit-product = {\,}]
  \pixel{pixel}
\usepackage{natbib}
\usepackage{widetext}
\usepackage{lscape}
\bibpunct{(}{)}{;}{a}{}{,} 
\usepackage{txfonts}
\usepackage[colorlinks]{hyperref}
\hypersetup{
colorlinks=true,
    citecolor={blue},
    linkcolor={blue}
}

\begin{document} 

\title{Testing solar surface flux transport models in the first days after active region emergence}
\titlerunning{Testing SFTM in the first days after active region emergence}
 
\author{N. Gottschling
          \inst{1}
          \and
          H. Schunker\inst{2}
          \and
          A. C. Birch\inst{1}
          \and
          R. Cameron\inst{1}
          \and
          L. Gizon\inst{1}\fnmsep\inst{3}
          }

 \institute{Max-Planck-Institut f\"ur Sonnensystemforschung, 
            Justus-von-Liebig-Weg 3, 37077 G\"ottingen, Germany\\
            \email{gottschling@mps.mpg.de}
        \and 
            School of Mathematical and Physical Sciences, University of Newcastle, Callaghan, New South Wales, Australia
        \and
            Institut f\"ur Astrophysik, Georg-August Universit\"at 
            G\"ottingen, 37077 G\"ottingen, Germany
 }

\date{}

\abstract
   {Active regions (ARs) play an important role in the magnetic dynamics of the Sun. Solar surface flux transport models (SFTMs) are used to describe the evolution of the radial magnetic field at the solar surface. The models are kinematic in the sense that the radial component of the magnetic field behaves like passively advected corks. There is however uncertainty about using these models in the early stage of active region evolution, where dynamic effects might be important.}
   {We aim to test the applicability of SFTMs in the first days after the emergence of active regions by comparing them with observations. The models we employ range from passive evolution to models where the inflows around active regions are included.}
   {We simulate the evolution of the surface magnetic field of 17 emerging active regions using a local surface flux transport simulation. The regions are selected such that they do not form fully-fledged sunspots that exhibit moat flows. The simulation includes diffusion and advection by a velocity field, for which we test different models. For the flow fields, we use observed flows from local correlation tracking of solar granulation, as well as parametrizations of the inflows around active regions based on the gradient of the magnetic field. To evaluate our simulations, we measure the cross correlation between the observed and the simulated magnetic field, as well as the total unsigned flux of the ARs, over time. We also test the validity of our simulations by varying the starting time relative to the emergence of flux.}
   {We find that the simulations using observed surface flows can reproduce the evolution of the observed magnetic flux. The effect of buffeting of the field by supergranulation can be described as a diffusion process. The SFTM is applicable after \SI{90}{\percent} of the peak total unsigned flux of the active region has emerged. Diffusivities in the range between $D=$~\SIrange{250}{720}{\kilo \meter \squared \per \second} are consistent with the evolution of the AR flux in the first five~days after this time. We find that the converging flows around emerging active regions are not important for the evolution of the total flux of the AR in these first five~days; their effect of increasing flux cancellation is balanced by the decrease of flux transport away from the AR.}
{}


\maketitle

\section{Introduction}

Active regions (hereafter ARs) are the surface signature of magnetic flux rising from the interior of the Sun. They are the site of eruptive events such as jets and flares, and play an important role in the solar dynamo.

During the emergence of active regions, their magnetic polarities move apart, and develop a tilt angle, with the leading polarity closer to the equator than the trailing polarity (e.g. \citealt{Schunker_2020}, for a review see \citealt{Driel_Gesztelyi_2015}). This is consistent with the footpoints of the flux being connected to the subsurface field, and separating due to the action of magnetic tension and drag force \citep{Chen_Rempel_2017, Schunker_2019}. \cite{Schunker_2019} calculated separation speeds of AR polarities and suggested that during emergence, the magnetic tension and drag force play a stronger role in transporting the magnetic field than diffusion. They also found that the scatter in polarity positions increases with time consistently with buffeting by supergranulation.

Studying the surface magnetic field of active regions helps to understand their evolution, and the buildup of poloidal field in the solar cycle. In addition to the systematic motions from e.g. magnetic tension, the processes on the solar surface that displace magnetic flux are the random motions of convective granulation and supergranulation on smaller scales as well as systematic flows on larger scales, such as differential rotation, the meridional flow, and inflows around active regions. The random convective motions can be treated as a diffusion process \citep{Leighton_1964}, which can be implemented in surface flux transport models (SFTMs) as a random walk. Estimates of the diffusion rate $D$ from observations typically indicate $D=\SI{250}{\kilo\meter\squared \per \second}$ \citep{Jafarzadeh_2014}, but higher values up to $D=\SI{500}{\kilo\meter\squared \per \second}$ have also been reported \citep{Wang_2002, Yeates_2020bipolar}.

Inflows on various spatial and temporal scales around evolved active regions have been measured by e.g. \citet{Gizon_2001, Haber_Hindman_2004, Komm_2012, Loeptien_2017, Braun_2019}. They span approximately \SI{10}{\degree} from the active region, and have velocities of about \SI{50}{\meter \per \second}. It is thought that the inflows may be driven by increased cooling in ARs \citep{Spruit_2003}. \citet{Cameron_2012} proposed these inflows as a possible mechanism for a non-linearity that regulates the solar cycle strengths in the solar dynamo. Recently, \citet{Gottschling_2021_Evolution} measured the evolution of the flows around emerging active regions from before to up to seven days after emergence, finding that the time between the AR emergence and the time at which inflows set in after the emergence increases with the total magnetic field of the AR. These inflows have velocities of about \SI{50}{\meter \per \second} as well, but appear to be smaller in extent than the inflows around evolved ARs. \citet{Gottschling_2021_Evolution} found no strong dependence of the amplitude of these inflows on the field strength of the active regions. The nature of these observed inflows in the first days after emergence is not clear, and their driving mechanism could be different from that of the inflows around evolved active regions. They could be the result of a passive emergence, in which the rising flux is affected by the supergranulation pattern \citep{Birch_2019}. Another scenario is that they are driven by the magnetic tension that moves the polarities apart in the first days after emergence, see \citet{Cameron_2010}, \citet{Schunker_2019}. In a three-dimensional MHD simulation of a rising flux tube in a rotating convection zone, \citet{Abbett_2001} also found converging flows.

Several studies have incorporated inflows around active regions in surface flux transport models \citep{DeRosa_Schrijver_2006, Jiang_2010, Cameron_2012, Yeates_2014, MartinBelda_2016, MartinBelda_2017_Inflows}. \citet{MartinBelda_2016} found that the inflows enhance flux cancellation, and can in conjunction with differential rotation produce a net tilt angle. This tilt is however too small compared with observed tilt angles. \citet{MartinBelda_2017_Inflows} investigated the effect on the large-scale field and found that the inflows can lead to a reduction of the axial dipole moment by \SI{30}{\percent}. Inclusion of the inflows into global SFTMs improve the match to the global dipole for the solar cycles 13 to 21 \citep{Cameron_2012}, and can account for the excess strength of the polar field at activity minimum in simulations, by effectively reducing the tilt angle of active regions \citep{Cameron_2010, Yeates_2014}. However, recently \citet{Yeates_2020bipolar} argued that this excess strength can be a result of the bipolar approximation of the active regions. On the other hand, \citet{Yeates_2014} found that the incorporation of the inflows (in form of a perturbation of the meridional flow) delays the dipole reversal times for solar cycle 23 with respect to the observed cycle.

The above studies used simple mathematical descriptions as parametrizations of the inflows. They were included either as a perturbation of the meridional flow at active latitudes \citep{Jiang_2010, Cameron_2012, Yeates_2014}, or as the gradient of the magnetic field \citep{DeRosa_Schrijver_2006}, with a normalization such that the extent and the amplitude of the inflows are similar to the observed values. The latter however raised the problem that flux of an isolated AR got 'trapped' by the inflow field due to the inward-directed flows from all sides, such that flux cannot escape the active region. The flux is pushed into small, long-lived clumps, which are not observed. Part of this effect might however be caused by the flux-dependent diffusion that was used \citep{MartinBelda_2016}.

In this work, we use a local surface flux transport model to simulate the evolution of the magnetic field of ARs, and compare it to the observed evolution. For this, we consider a sample of emerging active regions that take several days to cross the visible disk after the bulk of flux has emerged. The simulations include transport by diffusion and by advection due to surface flows. We test different models for both. For the advection, we use observed flow maps from correlation tracking of solar granules as well as flow parametrizations from the literature, motivated by the resemblance of the observed flows in the first days after emergence with the inflows around active regions.

This paper is structured as follows. In Sect.~\ref{sect_data}, we describe the sample of active regions on which we carry out the flux transport simulation. In Sect.~\ref{sect_simulation}, we describe the SFTM simulation that we use, as well as the different models of the flow field and the diffusion. Sect.~\ref{sect_results} presents the results, followed by a discussion (Sect.~\ref{sect_discussion}).

\section{Active region sample}
\label{sect_data}

\begin{figure}
\centering
\includegraphics[width=\hsize]{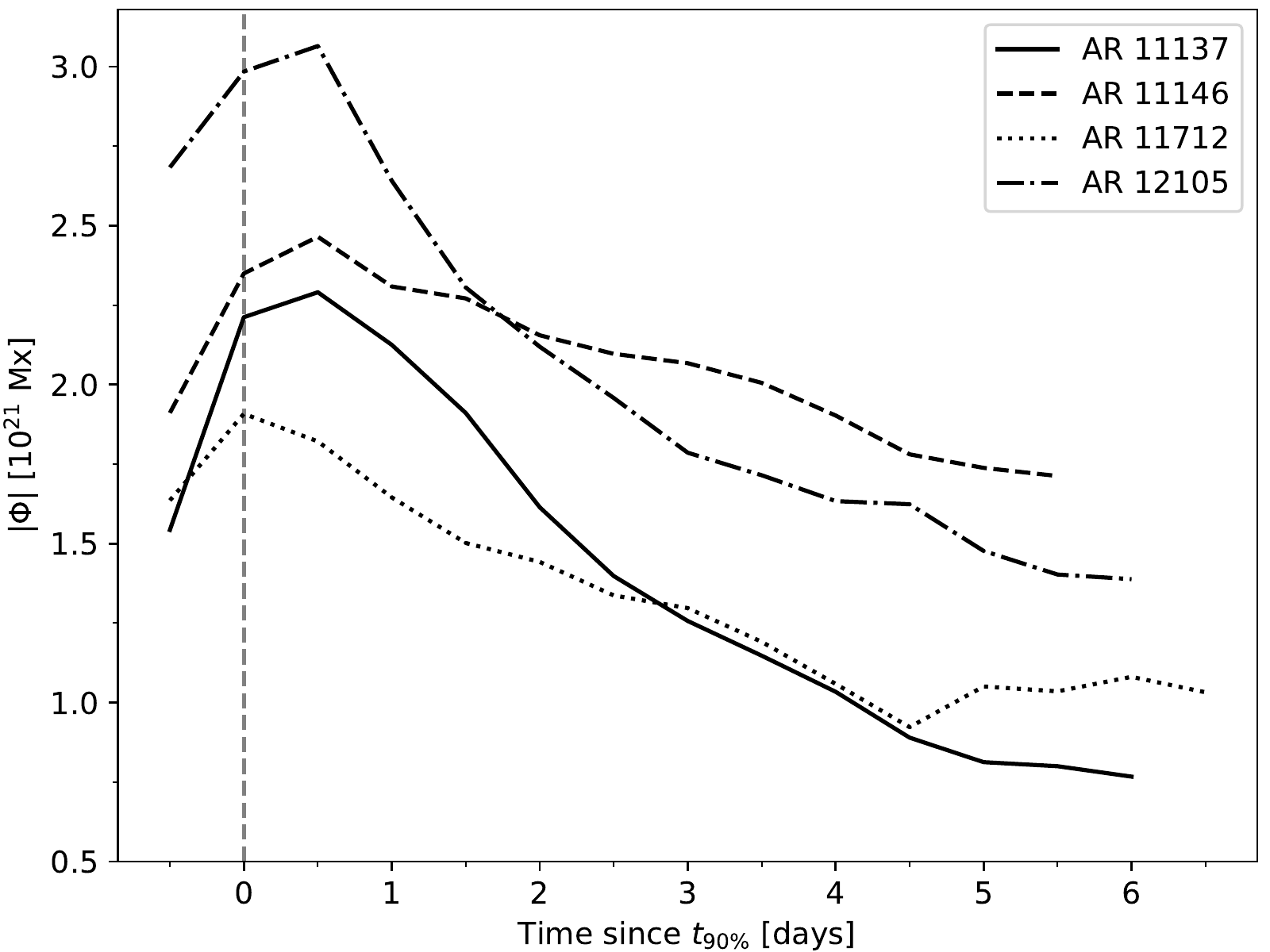}
  \caption[Examples of the evolution of the total unsigned magnetic flux]{Evolution of the total unsigned magnetic flux of four~ARs, calculated from twelve-hour averages of the $\mu$-corrected line-of-sight magnetograms. The ARs are aligned in time relative to the time $t_{\rm{90\%}}$ at which \SI{90}{\percent} of the maximum flux has emerged. In one of the shown examples, this coincides with the time of peak flux.}
      \label{plot_usflux_separation}
\end{figure}

We identify active regions that emerge into the quiet Sun and remain on the visible disk for multiple days after their emergence using the Solar Dynamics Observatory Helioseismic Emerging Active Regions (SDO/HEAR) survey \citep{Schunker_2016}. The survey consists of 182~emerging active regions that are observed between up to seven days before and after the time of emergence $t_0$, at which the region reaches \SI{10}{\percent} of the maximum total unsigned flux within the first \SI{36}{\hour} after first appearance in the NOAA record.

From the 182~active regions in the HEAR survey, we selected those regions that do not develop a fully-fledged sunspot with a clear penumbra over the disk passage. Fully-fledged spots show moat flow signatures \citep{Sheeley_1972}. At the grid scale that we use in our simulation, which is limited by the observed flows (see Sect.~\ref{subsect_flowmodels}), the moat flow spatially overlaps with the magnetic field of the spots. In the simulation, this would lead to a disruption of the spots into ring-like structures, which is inconsistent with observations. The sunspot identification is done with the sunspot quality number from \citet{Gottschling_2021_Evolution}, where a sunspot quality of 0 indicates a spot with a clear penumbra. After excluding these regions, 92~ARs are left in the sample.

For a comparison of the observed and the simulated magnetic field, the ARs have to remain on the visible disk for several days after the simulations are initialized. To select suitable ARs, we created data cubes of the line-of-sight magnetic field observed by the Helioseismic and Magnetic Imager onboard the Solar Dynamics Observatory (SDO/HMI, \citealt{Schou_2012}), projected to Plate Carree projection and corrected for the viewing angle $\mu$. The cubes have a field of view of \SI{60x60}{\degree} and a grid spacing of \SI{0.4}{\degree} in both longitude and latitude, centered on the active region centres as defined in the HEAR survey. The grid spacing is the same as that of the observed flow maps (Sect.~\ref{subsect_flowmodels}). On the temporal axis, the cubes span up to seven days before and after the time of emergence, in non-overlapping twelve-hour averages. We measured the total unsigned flux of each active region as the total unsigned flux in a disk of \SI{5}{\degree} radius around the center of the AR. From this, we determined the time $t_{\rm{90\%}}$ at which \SI{90}{\percent} of the maximum total unsigned flux of the active region has emerged, in the period covered by the HEAR survey. This is in analogy to the definition of the emergence time $t_0$. 

The observed flows have data coverage out to only \SI{60}{\degree} from disk center, which limits the last last time step $t_{\rm{last}}$ that can be used in the simulation. We identify $t_{\rm{last}}$ as the last time step where more than half of the field of view of the observed flow field is within a distance of \SI{60}{\degree} to the disk center.

For the sample selection, we require that the time between $t_{\rm{90\%}}$ and $t_{\rm{last}}$ is at least 5.5~days. This leaves 17~ARs. Appendix~\ref{appendix_ARlist} lists all ARs in this sample. Because of the selection criteria (the exclusion of ARs that form fully-fledged sunspots, and the requirement of several days of observations after most of the flux has emerged), the selected ARs are relatively weak and short-lived, and are well into their decaying phase at the end of the observations.

Fig.~\ref{plot_usflux_separation} shows a few examples of the evolution of the twelve-hour averaged total unsigned flux over time, relative to $t_{\rm{90\%}}$. In most cases, the peak flux occurs at the time $t_{\rm{90\%}}+0.5$~days. The average total unsigned flux over the sample of ARs at that time is \SI{1.65e21}{\Maxwell}, with a standard deviation of \SI{0.66e21}{\Maxwell}.

\section{Cork simulation for local surface flux transport}
\label{sect_simulation}

For the local flux transport simulation, we adapt the cork simulation of \citet{Langfellner_2018}. The simulation treats the magnetic field from an initial input magnetogram as individual, passive flux elements ('corks') in $x,y$ coordinates corresponding to the longitudinal and latitudinal axes of the projected input magnetic field map. At each simulation time step $\Delta t$, each cork moves a certain distance from its former position. There are two contributions to this motion: A diffusive part, which is realized as a random walk, and an advective part, which is realized as a flow field that moves each cork according to the velocity vector at its position.

\citet{Langfellner_2018} considered only unsigned magnetic field, and included spawning of randomly distributed new field as well as random removal of existing field. We do not include the random spawning and removal, as we are studying the AR polarities, for which the magnetic field is dominated by the emergence. We expand the simulation by considering positive and negative field and incorporating flux cancellation between the two. Corks of opposite polarity that move within \SI{1}{\mega \meter} of each other are removed from the simulation. The distance threshold is the same as that used by \citet{MartinBelda_2016}.

We initialize the simulations with individual time steps from the magnetogram cubes described in Sect.~\ref{sect_data}, recentered to the center of the active region at time $t_{\rm{90\%}}$. A magnetic flux density of \SI{1}{\Gauss} in the observations corresponds to one cork in the simulation. The output of the simulations are magnetic field maps for each simulation time step $\Delta t$. The (signed) magnetic flux density at each grid element in these maps is the difference between the number of positive and negative corks that have $x,y$ coordinates within that grid element.

We run the simulations with a simulation time step $\Delta t$ of 30~minutes. We average the simulation output to the same twelve-hour averages as the observations, for direct comparison between simulations and observations. For each AR, we run 20~realizations of each simulation model and average over them, to decrease the realization noise from the random-walk diffusion models. The number of realizations is limited by computation time. The results do however not differ from those with less (e.g. five) realizations.

\subsection{Flow models}
\label{subsect_flowmodels}

\begin{figure}
\centering
\includegraphics[width=\hsize]{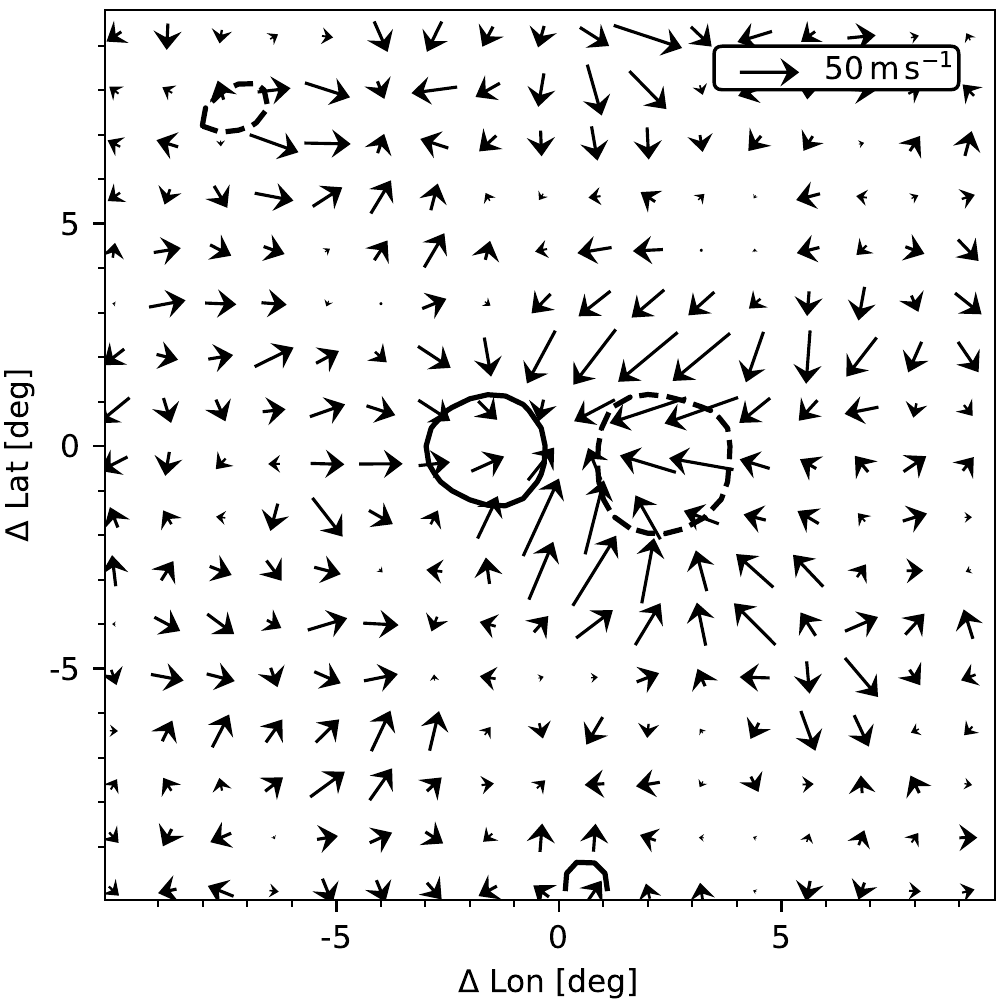}
  \caption[Average flows around a subsample of AR]{Average flows around the sample of ARs, in a twelve-hour average at four days after $t_{\rm{90\%}}$. The flow maps are smoothed with a Gaussian of width $\sigma=\SI{0.8}{\degree}$. The solid (dashed) contour indicates magnetic field at $+ (-) \SI{15}{\Gauss}$.}
      \label{plot_avobs_step}
\end{figure}

We use four different flow models in our simulations: no flow field ($\vec{u}=0$), flows from observations ($\vec{u}=\vec{u}_{\rm{obs}}$), and two parameterized, gradient-based models ($\vec{u}=\vec{u}_{\rm{\nabla B}}$ and $\vec{u}=\vec{u}_{\rm{\widetilde{\nabla B}}}$). We run each of these models with two different random-walk diffusion models. In an additional case, we add no diffusion. Thus, we have nine different simulation setups (see Table~\ref{table_simulationsetups}).

\begin{table}
\caption[Simulation setups]{Simulation cases. The simulations are run with a flow model (no flows, $\vec{u}=0$; observed flows, $\vec{u}_{\rm{obs}}$; parameterized inflows around active regions, $\vec{u}_{\rm{\nabla B}}$; or modified parameterized inflows, $\vec{u}_{\rm{\widetilde{\nabla B}}}$), and a diffusion model (no diffusion, $D=0$; constant diffusivity, $D_{\rm{c}}$; or flux-dependent diffusivity, $D_{\rm{f}}$). Fig.~\ref{plot_examplesim_vidframes_11137} shows snapshots from each simulation for AR~11137.}
\label{table_simulationsetups}
\centering 
\begin{tabular}{c c} 
\hline\hline
Flow model $\vec{u}$ & Diffusion model D\\ 
\hline
no flows ('\vec{u}=0') & $D_{\rm{c}}$ \\
no flows ('\vec{u}=0') & $D_{\rm{f}}$ \\
$\vec{u}_{\rm{obs}}$ & no diffusion ('D=0') \\
$\vec{u}_{\rm{obs}}$ & $D_{\rm{c}}$ \\
$\vec{u}_{\rm{obs}}$ & $D_{\rm{f}}$ \\
$\vec{u}_{\rm{\nabla B}}$ & $D_{\rm{c}}$ \\
$\vec{u}_{\rm{\nabla B}}$ & $D_{\rm{f}}$ \\
$\vec{u}_{\rm{\widetilde{\nabla B}}}$ & $D_{\rm{c}}$ \\
$\vec{u}_{\rm{\widetilde{\nabla B}}}$ & $D_{\rm{f}}$ \\
\hline
\end{tabular}
\end{table}

\textbf{No flows ($\vec{u}=0$)}: Here, we include no flow field in the simulation. Thus, only diffusion displaces the corks.

\textbf{Observed flows ($\vec{u}=\vec{u}_{\rm{obs}}$)}: In this model, we use the flow maps from \citet{Gottschling_2021_Evolution} for the ARs in the sample. The flows stem from local correlation tracking (LCT, \citealt{November_Simon_1988}) of solar continuum intensity images, and are based on the data processing by \citet{Loeptien_2017}, who used the Fourier local correlation tracking code \citep{Welsch_2004, Fisher_Welsch_2008} on full-disk continuum intensity images from SDO/HMI. Several changes in the data processing were made by \citet{Gottschling_2021_Evolution} in order to correct for additional systematic effects in the data from \citet{Loeptien_2017}. \citet{Gottschling_2021_Evolution} describe these changes in detail. The flow maps are in Plate Carree projection with a grid spacing of \SI{0.4}{\degree}, with the same centering and twelve-hour time steps as the magnetograms (see Sect.~\ref{sect_data}). In the SFTM, these are Fourier-interpolated to the \SI{30}{\minute} simulation time step $\Delta t$. Fig.~\ref{plot_avobs_step} shows the average magnetic field and flows over the sample of ARs, at four days after $t_{\rm{90\%}}$. Converging flows towards the center of the AR are visible, with velocities on the order of \SI{50}{\meter \per \second}. Inflows are more pronounced along the latitudinal axis than along the longitudinal axis.

\textbf{Parameterized inflows ($\vec{u}=\vec{u}_{\rm{\nabla B}}$)}: We adopt the parametrization of the inflows around active regions by \citet{DeRosa_Schrijver_2006}, who used it on the simulation by \citet{Schrijver_2001} to study the evolution of the field on large timescales, i.e. multiple rotations. Here, we apply it to the first few days after emergence, motivated by the resemblance of the observed flows to the inflows around active regions (see Fig.~\ref{plot_avobs_step}). The parametrization is
\begin{equation}
    \vec{U} = \alpha \nabla \left( |\Tilde{B}|^{\beta}\right), \label{eq_flowgradient}
\end{equation}
where $\vec{U}$ is the flow field $\left(\vec{u}_{\rm{lon}},\vec{u}_{\rm{lat}}\right)$, $\Tilde{B}$ is the magnetic field, smoothed with a Gaussian with a full width at half maximum (FWHM) of \SI{15}{\degree}, and $\alpha$ and $\beta$ are free parameters. \citet{MartinBelda_2017_Inflows} used Eq.~\ref{eq_flowgradient} with $\beta=1$ and normalized it such that the peak inflow velocity around an active region of \SI{10}{\degree} is \SI{50}{\meter \per \second}, with $\alpha=$ \SI{1.8e8}{\meter \squared \per \Gauss \per \second}. We adopt these choices for our model. The choice of $\beta$ is further motivated by the observed flux density corresponding to the product of the field strength and the filling factor. If the inflows are driven by excess cooling, as suggested by \citet{Spruit_2003}, they are proportional to the filling factor, and thus $\beta=1$. Because we investigate active regions shortly after their emergence, their extent is considerably smaller than \SI{10}{\degree}. Therefore, the inflows in this model will have too large extents and amplitudes well below \SI{50}{\meter \per \second}. In previous studies, these parameterized inflows led to flux clumping. In Appendix \ref{appendix_flowfieldbalancingdiffusion}, we therefore compare the inflows from this parametrization to the flow field that would balance the diffusion of a flux concentration. If the inflows were stronger than this flow field, they would lead to flux clumping. We find that the inflows from this parametrization are too weak to cause flux clumping, as was also motivated above.

\textbf{Modified parameterized inflows ($\vec{u}=\vec{u}_{\rm{\widetilde{\nabla B}}}$)}: With this model, we aim for a parametrization of the inflows that more closely resembles the observed flows on our sample of comparatively small (and young) active regions, rather than the evolved large active regions on which the parameters of the above model ($\vec{u}_{\rm{\nabla B}}$) are based. In order to capture both the spatial extent of the flows as well as their amplitude, we compared the extent of the magnetic field, smoothed with different levels of spatial smoothing, to the observed flow field. We found that the magnetic field smoothed with a Gaussian of $\sigma= \SI{2}{\degree}$ has a similar extent as the observed flows. The observed average inflow velocities increase from about \SI{10}{\meter \per \second} at $t_{\rm{90\%}}$ to \SIrange{40}{50}{\meter \per \second} at $t_{\rm{90\%}}+5$~days. To capture this evolution, we fit a line between the observed flow velocities and the gradient of the smoothed magnetic field at the same location, for each twelve-hour time step. The fit considers the area of \SI{20x20}{\degree} around the center of the AR. We further selected only those pixels that lie within \SI{2}{\degree} of an absolute magnetic field density above \SI{20}{\Gauss}. We then fit a line to the relation between the flow velocity and the gradient against time. We use the slope and intercept of this fit to calculate the normalization $\alpha$ in Eq.~\ref{eq_flowgradient} for each time step.

\subsection{Diffusion models}
\label{subsect_diffusionmodels}

We consider the cases where diffusion is the same for all flux elements (${D_{\rm{c}}}$), where it is flux-dependent ($D_{\rm{f}}$), and without diffusion ($D=0$).

\textbf{Constant diffusivity ($\mathbf{D_{\rm{c}}}$)}: In this model, the step length of the random walk of each cork along each axis is drawn from a normal distribution, with a standard deviation corresponding to a diffusion rate $D$ that is constant for all corks. We run simulations with diffusivities in the range between $D=250$ and~ \SI{722.5}{\kilo\meter\squared\per\second}.

\textbf{Flux-dependent diffusivity ($\mathbf{D_{\rm{f}}}$)}: This model is based on \citet{DeRosa_Schrijver_2006}. In their simulation, they treated the magnetic field as flux concentrations that can contain a varying amount of flux and that can merge and break up. The random walk step length $\Delta r$ of a flux concentration depends on the amount of flux in the concentration as
\begin{eqnarray}
    \Delta r &=& C(|\Phi|)\sqrt{4D\Delta t}\label{eq_DeRS_deltar}\\
    C(|\Phi|) &=& 1.7\ \rm{exp} \left( \frac{-|\Phi|}{3\times 10^{19}\ \rm{Mx}} \right)\label{eq_DeRS_C},
\end{eqnarray}
where $|\Phi|$ is the absolute flux within a flux concentration and $D$ is the diffusion rate \citep{Schrijver_2001}. This treatment of the magnetic field is different from our simulation, where we consider individual corks that have a constant amount of flux and perform individual random walks. To implement a flux-dependent diffusion rate in our simulation comparable to \citet{DeRosa_Schrijver_2006}, at each simulation time step we calculate the cork density on a grid in longitude and latitude with grid spacing \SI{0.4}{\degree}, oversample it 4x4 pixels and smooth it by \SI{0.1}{\degree}. For each cork, we then calculate the width $\Delta r$ of the normal distribution from which the random walk step length is drawn using the cork density at the corks' position, with Eqs.~\ref{eq_DeRS_deltar} and \ref{eq_DeRS_C}. As reference diffusivity $D$ in Eq.~\ref{eq_DeRS_deltar}, we use $D=\SI{250}{\kilo\meter\squared \per \second}$.

\textbf{No diffusion ($\mathbf{D=0}$)}: For one simulation using the observed flows, we add no diffusion. In this case, only the flow field displaces the magnetic field.

\section{Evaluation of the models}
\label{sect_results}

We run simulations using the four different inflow models (Sect.~\ref{subsect_flowmodels}) and the three different diffusion models (Sect.~\ref{subsect_diffusionmodels}), on the sample of 17~active regions (Sect.~\ref{sect_data}). To evaluate the different simulations in comparison to the observations, we measure for each time step the total unsigned flux within a disk of the central \SI{5}{\degree}, as well as the cross correlation between the observations and the simulations. We then study different simulation start times.

\subsection{Active region flux as a function of time}
\label{subsect_flux}

\begin{figure*}
\centering
\includegraphics[width=\hsize]{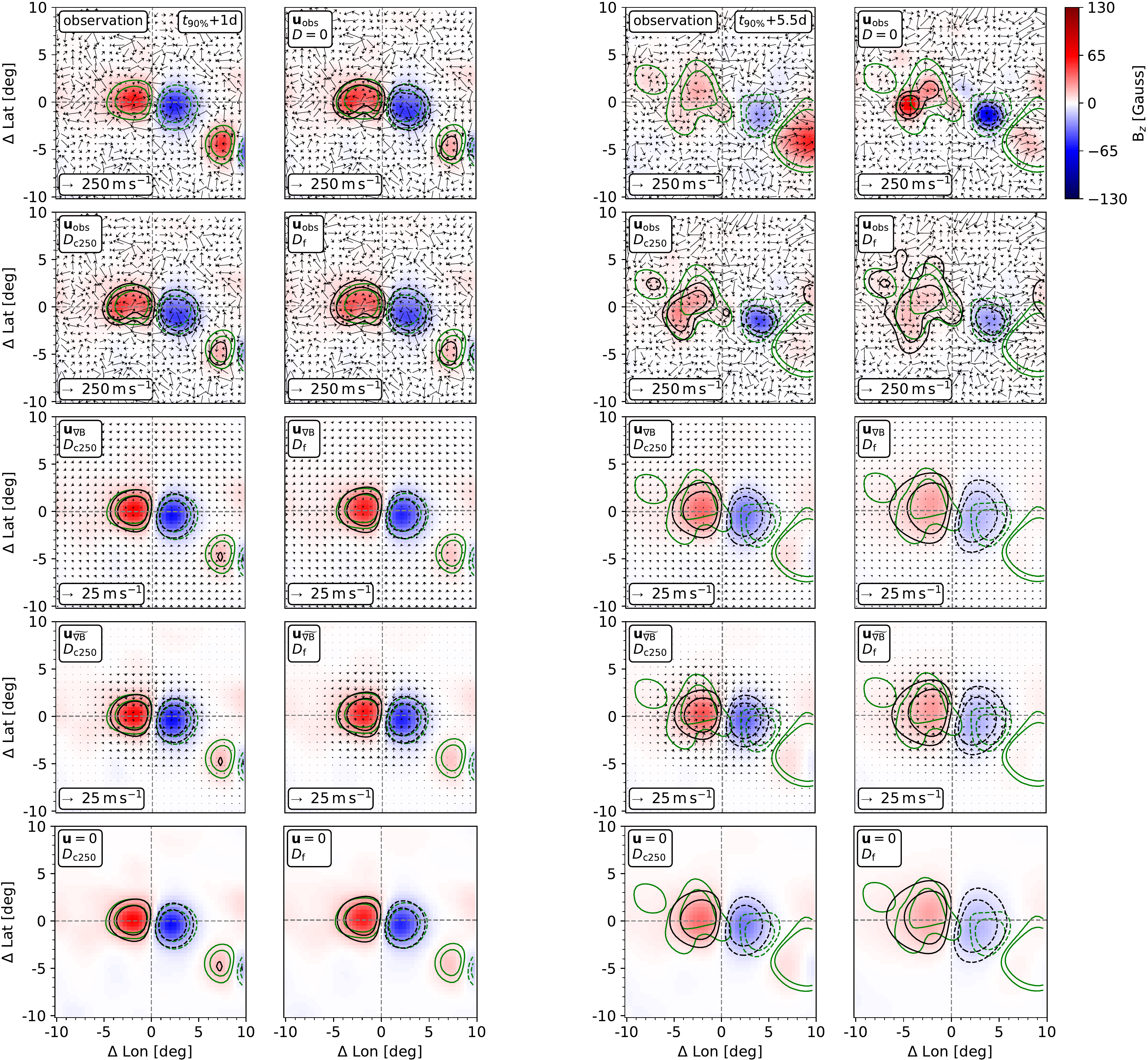}
  \caption[Simulation example time steps for AR~11137]{Two example time steps of the observations and the simulations, for AR~11137. The two columns on the left show the first time step after simulation start, that is, 0.5~days after $t_{\rm start} = t_{\rm{90\%}}+0.5$~days. The two columns on the right show the time step at the end of the simulations, at $t_{\rm{90\%}}+5.5$~days. The times are indicated in the upper right corner of the top left panels. At each of the two time steps, the top row shows the observed magnetic field and flows (left) and the simulation using observed flows and no additional diffusion (right). For all other rows, the left (right) panels show simulations with constant (flux-dependent) diffusivity $D_{\rm{c}}$ ($D_{\rm{f}}$). From second row to bottom: observed flows, flows according to the parameterized inflow model, flows according to the modified parameterized inflow model, and no flows. The diffusion in the cases of constant diffusivity is \SI{250}{\kilo\meter\squared \per \second}. The arrows indicate the observed and the parameterized flows, for the respective observations and simulations. Reference arrows are given in the lower left corners of each panel. Red (blue) indicates positive (negative) radial magnetic field. All maps have the same saturation, at $\pm$ the rounded maximum absolute field strength in the central \SI{10}{\degree} from all simulation time steps.
  The green (black) contours indicate levels of half and quarter of the minimum and maximum magnetic field in the central \SI{10}{\degree} of the observation (of each simulation), for each time step individually.}
  \label{plot_examplesim_vidframes_11137}
\end{figure*}

\begin{figure*}
\centering
\includegraphics[width=\hsize]{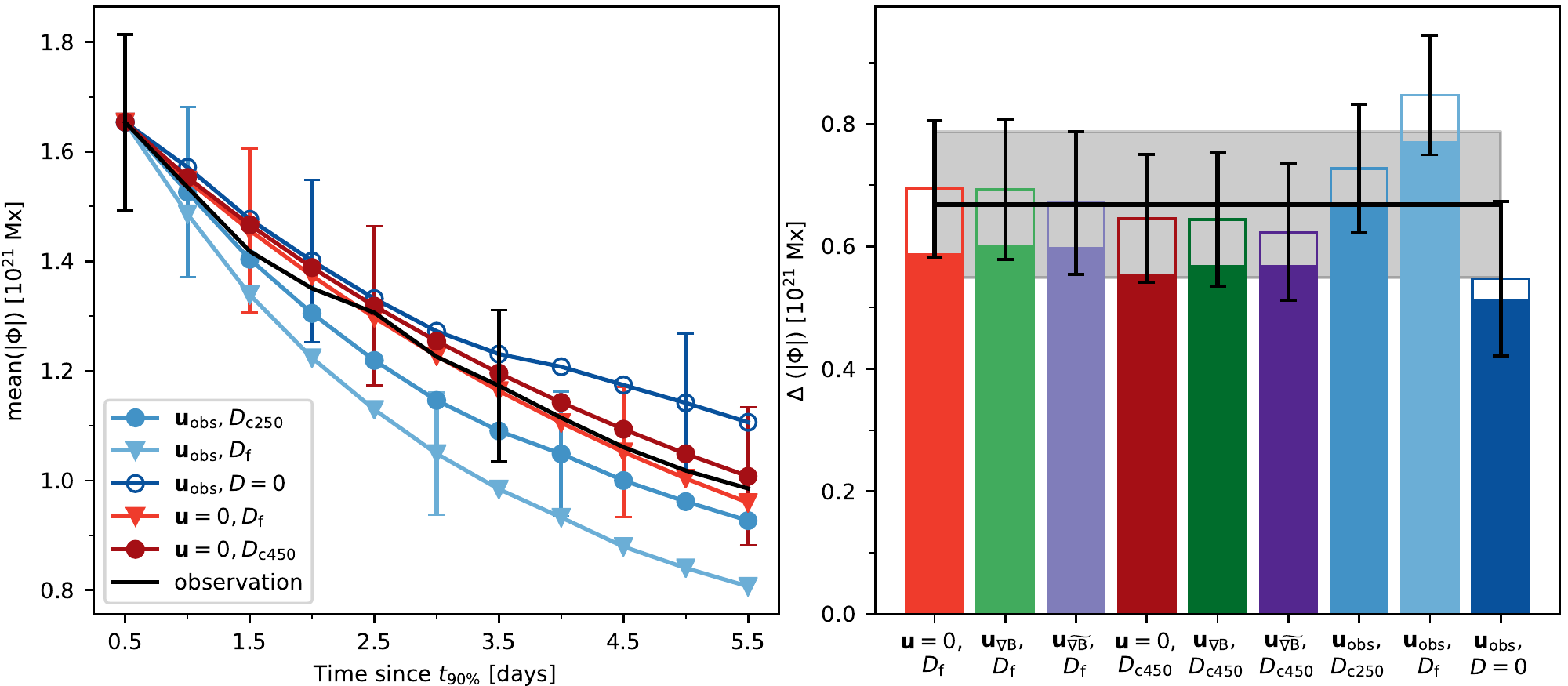}
  \caption[Evolution of AR flux in observations and simulations, at D=\SI{450}{\kilo\meter\squared\per\second}]{Left: Evolution of the total unsigned flux over the central disk with radius \SI{5}{\degree}, averaged over the sample of active regions, for the observations and some of the simulations. The error bars indicate the standard error over the sample. Only every sixth error bar is plotted, for readability. The data point at $t_{\rm{90\%}}+0.5$~days is the initial condition of the simulations. The diffusion in the simulation with constant diffusivity and no flows is $\SI{450}{\kilo \meter \squared \per \second}$. Right: The amount of flux loss between the last time step of the simulations, at $t_{\rm{90\%}}+5.5$~days, and the time when the simulations are initialized, at $t_{\rm{90\%}}+0.5$~days. The black line indicates the flux loss in the observations over the same period, the gray shaded area indicates the standard error. The solid and contoured bars indicate flux loss due to cancellation and advection, respectively.}
     \label{plot_Uflux_t90p1_sm2_vs_time_total_balance}
\end{figure*}

\begin{figure}
\centering
\includegraphics[width=\hsize]{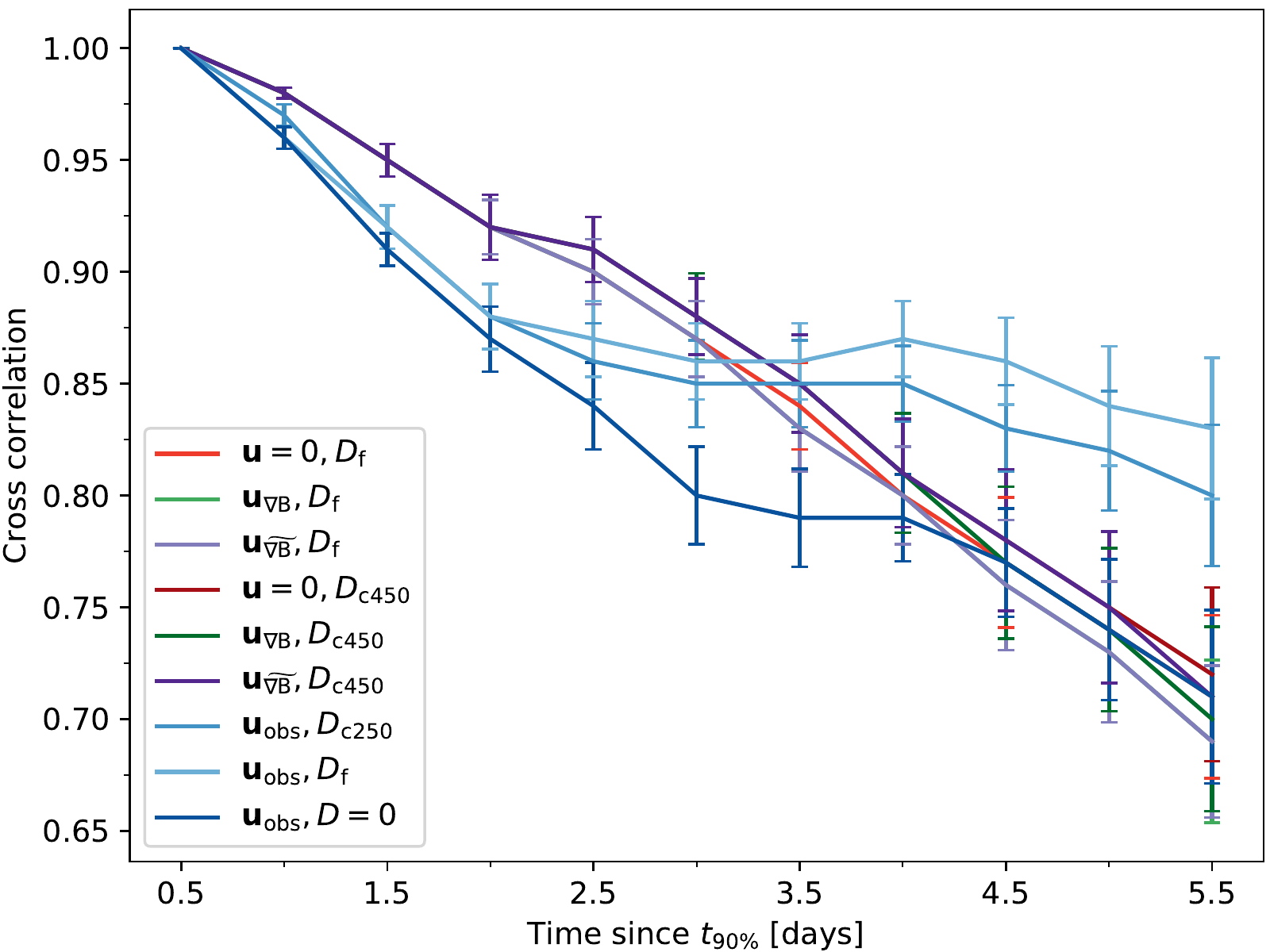}
  \caption[Evolution of cross correlations between observations and simulations]{Evolution of the average cross correlation between the observed field and the simulations. The error bars indicate the standard error over the sample. The data point at $t_{\rm{90\%}}+0.5$~days is the initial condition of the simulations.}
  \label{plot_CC_vs_time_t90p1}
\end{figure}

In this section, we initialize the simulations of each AR with the observed magnetic field at the time $t_{\rm{start} }= t_{\rm{90\%}}+0.5$~days. In the simulations that use the observed flow field ($\vec{u}_{\rm{obs}}$), stagnation points at which the velocities are zero have an infinitely small width. Magnetic field therefore tends to accumulate in very confined spaces. To reduce the emphasis on these small-scale features, we smooth the simulated magnetic field maps with a Gaussian with a width of $\sigma = \SI{0.8}{\degree}$. A broader smoothing results in higher cross correlations, as the small-scale structures are smeared out.

Fig.~\ref{plot_examplesim_vidframes_11137} shows the magnetic field from the observations and the different simulations, for AR~11137, at the beginning and at the end of the simulations. For reference, Fig.~\ref{plot_examplesim_11137} shows all time steps for four of the simulations along with the observations. In the observations, the leading polarity moves in the prograde direction and towards the equator over time. The trailing polarity is deformed. Its field disperses, and clumps of flux leave the flux concentration. Both polarities loose flux, due to flux cancellation between the two polarities as well as advection away from the active region.

In the simulations using the observed flows, the motions of the polarities resemble those in the observations (cf. the green and black contours in first and second row in Fig.~\ref{plot_examplesim_vidframes_11137}, and second row from top in Fig.~\ref{plot_examplesim_11137}). This indicates that the supergranulation, which is the dominant signal in the flow maps, is effective in buffeting the magnetic field polarities. Fig.~\ref{plot_examplesim_vidframes_11137} also shows the differences in the flow fields of the different models: The parameterized inflow model, $\vec{u}_{\rm{\nabla B}}$, yields extended flows with low velocities on the order of \SI{5}{\meter \per \second} (up to about \SI{10}{\meter \per \second} for larger ARs in the sample). The modified parameterized inflow model, $\vec{u}_{\rm{\widetilde{\nabla B}}}$, yields more confined flows with higher velocities on the order of \SI{20}{\meter \per \second} (up to about \SI{40}{\meter \per \second} for larger ARs in the sample), which is similar to the observed inflows (cf. Fig.~\ref{plot_avobs_step}). Fig.~\ref{plot_examplesim_vidframes_average} shows the averaged magnetic field and flows from all AR simulations and observations.

The flux-dependent diffusivity $D_{\rm{f}}$ (cf. Eq.~\ref{eq_DeRS_C}) increases the random walk step lengths of the magnetic flux elements (corks) in concentrations of weaker field, and decreases it for corks in concentrations of stronger field. The active regions considered here have relatively low magnetic field strengths (typically below \SI{100}{\Gauss}). Therefore, the flux-dependent diffusivity increases the diffusivity for most of the flux elements in the simulation, and decreases it only for a small amount of flux at the center of some AR polarities. For comparison, we run simulations with a range of constant diffusivities, from $D_{\rm{c250}} = \SI{250}{\kilo\meter\squared\per\second}$ which corresponds to the reference value of $D_{\rm{f}}$, to $D_{\rm{c722.5}}=\SI{722.5}{\kilo\meter\squared\per\second}$, in increments of \SI{50}{\kilo\meter\squared\per\second}. $D_{\rm{c722.5}}$ corresponds to each cork experiencing no surrounding magnetic flux in the flux-dependent model $D_{\rm{f}}$.

To evaluate how well the simulations reproduce the evolution of AR flux in the observations, we calculate for each time step the total unsigned flux within a disk of the central \SI{5}{\degree}, in the simulations and in the observations. The left panel of Fig.~\ref{plot_Uflux_t90p1_sm2_vs_time_total_balance} shows the average total unsigned flux of some of the simulations as well as the observations as a function of time, calculated from all regions in the sample, for $D_{\rm{c450}}=\SI{450}{\kilo\meter\squared\per\second}$. Appendix~\ref{appendix_diffusivities} shows the cases of $D_{\rm{c250}}$ and $D_{\rm{c722.5}}$. The results for ($\vec{u}_{\rm{\nabla B}}, D_{\rm{f}}$), ($\vec{u}_{\rm{\widetilde{\nabla B}}}, D_{\rm{f}}$), ($\vec{u}_{\rm{\nabla B}}, D_{\rm{c450}}$), and ($\vec{u}_{\rm{\widetilde{\nabla B}}}, D_{\rm{c450}}$) are very similar to the corresponding cases of ($\vec{u}=0, D_{\rm{f}}$) and ($\vec{u}=0, D_{\rm{c450}}$), and are therefore not shown in the left panel of Fig.~\ref{plot_Uflux_t90p1_sm2_vs_time_total_balance}.

Flux loss in active regions is the result of cancellation of opposite polarity flux as well as advection of flux away from the region. With our simulation, we can measure how much flux is lost in each of these processes separately. The change in total unsigned flux within an area from one time step to the next can be written as:
\begin{equation}
    \Phi(t+\Delta t) = \Phi(t) + \Phi_{\rm{a\_in}} - \Phi_{\rm{a\_out}} - \Phi_{\rm{c}},
\end{equation}
where $\Phi(t)$ is the total unsigned flux in the area at time $t$, $\Delta t$ is the simulation time step of \SI{30}{\minute}, $\Phi_{\rm{a\_in}}$ is the flux advected into the area, $\Phi_{\rm{a\_out}}$ is the flux advected out of it, and $\Phi_{\rm{c}}$ is the flux lost due to cancellation of opposite-polarity field. In our simulation, this corresponds to counting all corks that move into or out of the AR area or cancel with an opposite-polarity cork. As the AR area, we use the central disk with radius \SI{5}{\degree}.

The right panel of Fig.~\ref{plot_Uflux_t90p1_sm2_vs_time_total_balance} shows the total amount of flux lost between the starting time $t_{\rm{start} } = t_{\rm{90\%}}+0.5$~days and five days later, for all nine simulations. The black line indicates the flux loss in the observations over the same period. Appendix~\ref{appendix_diffusivities} shows the simulations with $D_{\rm{c250}}$ and $D_{\rm{c722.5}}$, which are within the error consistent with the observations as well. The diffusivities that are consistent with the observed flux loss therefore range from about \SIrange{250}{720}{\kilo\meter\squared\per\second}. In Appendix~\ref{appendix_analyticalsolultionDc}, we derive an analytical solution of the flux loss due to cancellation between two diffusing Gaussian distributions, in the case of no flow field and constant diffusion. This is in good agreement with the flux loss due to cancellation in the corresponding simulation. With this small sample of ARs, and a time period of only 5~days, we cannot constrain the diffusivity further.

Fig.~\ref{plot_Uflux_t90p1_sm2_vs_time_total_balance} also shows that the total amount of flux loss is very similar for all models that use the same diffusion model ($D_{\rm{f}}$ or $D_{\rm{c}}$). Increasing the strength of the inflows (no inflows $\vec{u}=0$, weak inflows $\vec{u}_{\rm{\nabla B}}$, stronger inflows $\vec{u}_{\rm{\widetilde{\nabla B}}}$) leads to both more flux loss due to flux cancellation and less flux loss due to advection, for both diffusion models. These two effects partially cancel out, such that the net effect of the parameterized inflow models on the evolution of AR flux is small. In conclusion, the inflows are not important in the evolution of the flux budget of the AR in the first five days, but might play a role in the distribution of the surrounding field.

In all three simulations that use the observed flows, the magnetic flux loss is consistent with that in the observations. The two models that include additional diffusion (blue triangles and filled circles) loose more flux than the simulation that uses the observed flow field and no additional diffusion (blue circles). This is because the additional random walk diffusion adds to the diffusion from the supergranulation and therefore enhances cancellation. We conclude that the bulk transport from supergranulation provides a means by which flux is carried away from the active region, which is consistent with a diffusion process.

\subsection{Cross correlation as a function of time}
\label{subsect_crosscorrelation}

As a second evaluation of the simulations, we calculate the cross correlation between the observed and the simulated magnetic field, in a window of \SI{10x10}{\degree} around the center of each map for each of the 17~ARs. The window size is chosen to exclude other active regions in the field of view, which emerge at a later time or are significantly larger than the target active region, such that they exhibit moat flows (see for example the lower right corner in the top left panel of Fig.~\ref{plot_examplesim_vidframes_11137}) As in Sect.~\ref{subsect_flux}, we use the simulations that are initialized at $t_{\rm{90\%}}+0.5$~days.

Fig.~\ref{plot_CC_vs_time_t90p1} shows the average cross correlation of the simulations with the observations as a function of time. The cross correlation decreases monotonically for all simulations that use no or parameterized flows. The differences between the simulations with the same diffusion model and different flow models indicate that the small-scale distribution of the field is different. Comparing the simulations with the same flow model and different diffusion models, the cases with flux-dependent diffusivity $D_{\rm{f}}$ have higher cross correlations to the observations than the $D_{\rm{c450}}$ cases. This is within the error bars, however. The simulation using the observed flow field and no additional diffusion ($\vec{u}_{\rm{obs}}, D=0$) has a lower correlation to the observations than the other models for most of the four~days of simulation time, whereas the two simulations using the observed flow field and additional diffusion remain at a constant cross correlation (within errors) from 2-3~days onward. In the last few time steps, the cross correlation of these is larger than for cases with no or parameterized flows. The low cross correlation in the $\vec{u}_{\rm{obs}}, D=0$ case is a result of flux being dragged by the flow into a different supergranular downflow lane in the simulation than in the observation. The additional diffusion in the cases of ($\vec{u}_{\rm{obs}}, D_{\rm{c450}}$) and ($\vec{u}_{\rm{obs}}, D_{\rm{f}}$) mitigates this, which therefore have a higher cross correlation.

\subsection{Changing the simulation start time}
\label{subsect_params_t90}

\begin{figure}
\centering
\includegraphics[width=\hsize]{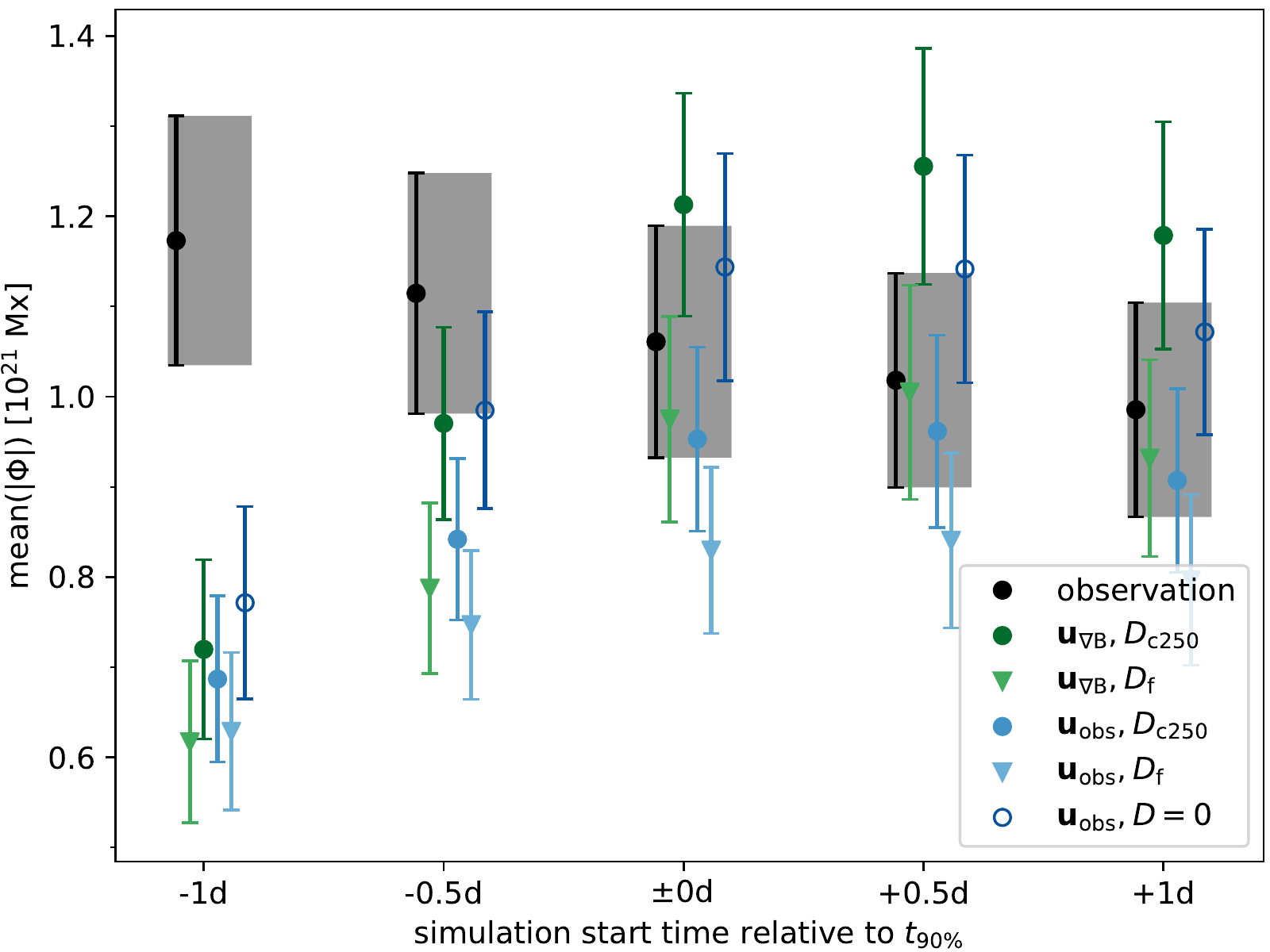}
  \caption[Total unsigned flux for different simulation start times]{Total unsigned flux for different simulation start times relative to $t_{\rm{90\%}}$, at 4.5~days after each simulation start time. The diffusion in the case of constant diffusivity is $\SI{250}{\kilo \meter \squared \per \second}$.}
 \label{plot_Uflux_vs_t90_fxtstep}
\end{figure}

In Sects.~\ref{subsect_flux} and~\ref{subsect_crosscorrelation}, we studied simulations initialized at $t_{\rm{90\%}}+0.5$~days. Here, we examine the dependence on the initial condition of the simulation, by initializing the simulation with the magnetic field at times $t_{\rm{90\%}}-1$~day, $t_{\rm{90\%}}-0.5$~days, $t_{\rm{90\%}}$, and $t_{\rm{90\%}}+1$~day.

Fig.~\ref{plot_Uflux_vs_t90_fxtstep} shows the total unsigned flux for the simulations initialized at the five different starting times, at 4.5~days after the initial condition of each simulation (that is, the same time has elapsed for all simulations). The simulations starting at $t_{\rm{90\%}}+0.5$~days yield the highest fluxes, because this time most often coincides with the time of peak flux of the active regions. Starting at later and earlier times decreases the AR flux in the simulations. In the two cases starting before $t_{\rm{90\%}}$, the simulated fields differ largely from the observations, as most of the flux has not yet emerged onto the surface.

\section{Discussion}
\label{sect_discussion}

In this work, we compared the observed evolution of 17~emerging active regions with surface flux transport simulations of these regions. We considered nine types of simulations with different models for diffusion and surface flows. We used models where the diffusion is zero, where it is the same for all flux elements, and where it is flux-dependent. For the surface flows, we used observations from local correlation tracking, parameterized models of the inflows around active regions, as well as no flows. We compared the evolution of the magnetic field in the observations and the simulations by calculating the cross correlation as well as the total unsigned flux of the ARs for all time steps. In addition, we tested the validity of the transport simulation for the study of young ARs by varying the starting time of the simulations relative to the time when the bulk of the AR flux has emerged.

We find that simulations using the observed flows can describe the evolution of the total unsigned flux of the ARs starting from the time when \SI{90}{\percent} of the AR flux has emerged. The supergranular motions act as a random walk process in buffeting the magnetic field polarities. This finding from our simulation complements the observations by \citet{Schunker_2019}, who measured the standard deviation of the positions of AR polarities and draw the same conclusion. However, from our simulation we cannot make a statement whether the buffeting is flux-dependent or not. Additional diffusion improves the small-scale structure (measured as the cross correlation). Our findings allow for diffusion rates from the supergranular motions between \SIrange{250}{720}{\kilo \meter \squared \per \second}. The large range is due to the small sample of active regions which was suited for this study, as well as the limitation to about five~days for the simulations.

The converging flows around emerging active regions, which we included as parameterized models, increase flux cancellation in the AR in the first five~days after \SI{90}{\percent} of the AR peak total unsigned flux have emerged. The resulting decrease in total unsigned flux is balanced by the decreased advection away from the AR, such that the evolution of the total flux associated with the AR is similar in the different models.

\begin{acknowledgements}
N.G. is a member of the International Max Planck Research School (IMPRS) for Solar System Science at the University of Göttingen.
N.G. conducted the data analysis, contributed to the interpretation of the results, and wrote the manuscript.
The HMI data used here are courtesy of NASA/SDO and the HMI Science Team.
We acknowledge partial support from the European Research Council Synergy Grant WHOLE SUN \#810218.
The data were processed at the German Data Center for SDO, funded by the German Aerospace Center under grant DLR 50OL1701.
This research made use of Astropy,\footnote{http://www.astropy.org} a community-developed core Python package for Astronomy \citep{astropy:2013, astropy:2018}. This work used the NumPy \citep{oliphant2006guide}, SciPy \citep{2020SciPy-NMeth}, pandas \citep{mckinney-proc-scipy-2010} and Matplotlib \citep{Hunter:2007} Python packages.

\end{acknowledgements}


\bibliographystyle{aa.bst}
\bibliography{Jabref_PaperII.bib}


\begin{appendix}

\section{List of ARs used in the simulation}
\label{appendix_ARlist}
The NOAA numbers of active regions used in the flux transport simulations:
11088, 11137, 11145, 11146, 11167, 11288, 11437, 11547, 11624, 11626, 11712, 11786, 11789, 11811, 11932, 12064, 12105.

\section{Flow field balancing diffusion}
\label{appendix_flowfieldbalancingdiffusion}

\begin{figure}
\centering
\includegraphics[width=\hsize]{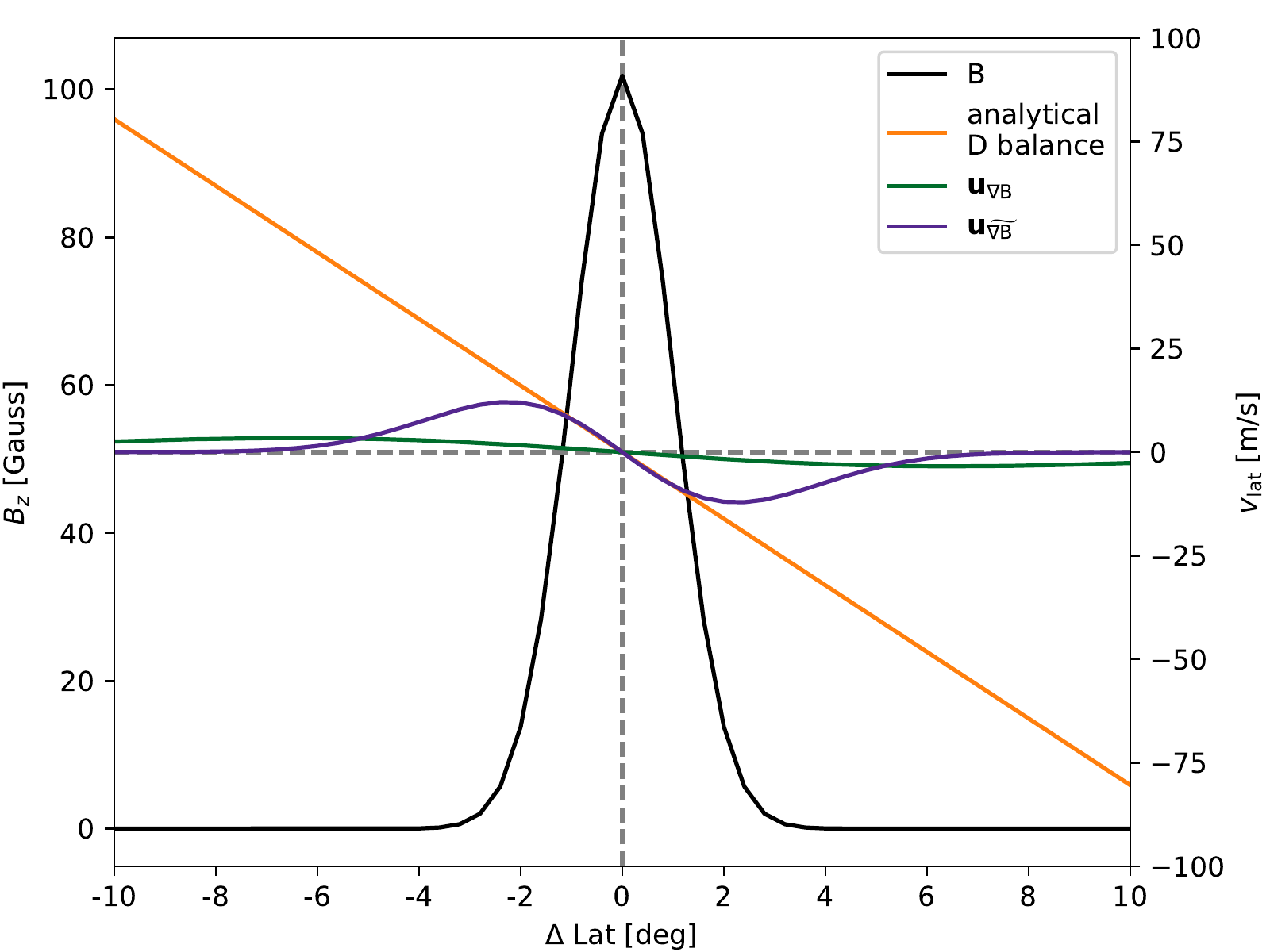}
  \caption[Flow profiles around artificial data]{Latitudinal cut through artificial data. The black line indicates the magnetic field, modeled as a Gaussian with a FWHM of \SI{2.4}{\degree}. The green line indicates the inflow model of $\vec{u}_{\rm{\nabla B}}$ for the shown magnetic field. The purple line indicates the inflow model of $\vec{u}_{\rm{\widetilde{\nabla B}}}$, at 2~days after $t_{\rm{90\%}}+0.5$~days. The orange line shows the inflow profile that compensates the diffusion of the magnetic field, for a constant diffusivity of \SI{250}{\kilo \meter \squared \per\second}.}
      \label{plot_Dcomp}
\end{figure}

We want to analyze the influence of the model inflows on the evolution of the magnetic field. For this, we calculated the flow field that compensates the diffusion of a flux distribution, and compared it with the model inflows. For the magnetic field, we considered a 2D Gaussian distribution:
\begin{equation}
    B(x,x_0,y,y_0,\sigma) = \frac{1}{2\pi \sigma^2} \mathrm{exp} \Bigg\{ - \left(\frac{(x -x_0)^2}{2 \sigma^2} + \frac{(y -y_0)^2}{2 \sigma^2} \right) \Bigg\}. \label{Gaussian2D}
\end{equation}
Its evolution in the presence of advection and diffusion is governed by the advection-diffusion equation (see for example \citealt{Leighton_1964}):
\begin{equation}
 \frac{\partial B} {\partial t} + \nabla \cdot (\vec{u} B)= D \nabla^2 B, \label{eqconvdiff_full}
\end{equation}
where $B$ is the radial magnetic field, $\vec{u}$ is the flow field, and $D$ is the diffusivity. In the situation where the advection and the diffusion balance each other, such that $\frac{\partial B}{\partial t}$ = 0, Eq.~\ref{eqconvdiff_full} reduces to
\begin{equation}
    \nabla \cdot (\vec{u} B) = D \nabla^2 B. \label{eqsteadystate}
\end{equation}
\pagebreak
Using the vector identity
\begin{equation}
    \nabla \cdot (\vec{u} B) = B(\nabla \cdot \vec{u}) + (\nabla B) \cdot \vec{u},
\end{equation}
Eq.~\ref{eqsteadystate} then gives
\begin{equation}
    B(\nabla \cdot \vec{u}) + (\nabla B) \cdot \vec{u} = D \nabla^2 B.
\end{equation}
Using Eq.~\ref{Gaussian2D} and its derivatives gives
\begin{eqnarray}
    \left[ \frac{\partial u_x}{\partial x} + \frac{\partial u_y}{\partial y} - \frac{(x-x_0)}{\sigma^2}u_x - \frac{(y-y_0)}{\sigma^2}u_y \right] = \nonumber \\
    \left[ \frac{-2D}{\sigma^2} + \frac{D}{\sigma^4} \left( (x-x_0)^2+ (y-y_0)^2 \right) \right].
\end{eqnarray}
From this we find the solution as
\begin{equation}
    \vec{u} = - \frac{D}{\sigma^2} 
    \begin{pmatrix}
    (x-x_0) \\ (y-y_0) \label{eq_velcomp_result}
    \end{pmatrix}.
\end{equation}
Fig.~\ref{plot_Dcomp} shows a latitudinal cut through the center of a 2D Gaussian with a full width at half maximum of \SI{2.4}{\degree} and a peak field strength of about \SI{100}{\Gauss}, along with the corresponding parameterized inflow models (green and purple lines) and the flow field which balances the effect of diffusion, from Eq.~\ref{eq_velcomp_result}. The figure shows that the $\vec{u}_{\rm{\nabla B}}$ inflow model has velocities that are too low to compensate the diffusion. The $\vec{u}_{\rm{\widetilde{\nabla B}}}$ model is similar to the diffusion within \SI{\pm2}{\degree} of the center of the field distribution, and in fact can overcome the diffusion, leading to flux clumping. 

\section{Evolution of observations and example simulations for AR~11137}

\begin{landscape}

\begin{figure}
\centering
\includegraphics[height=0.58\textheight]{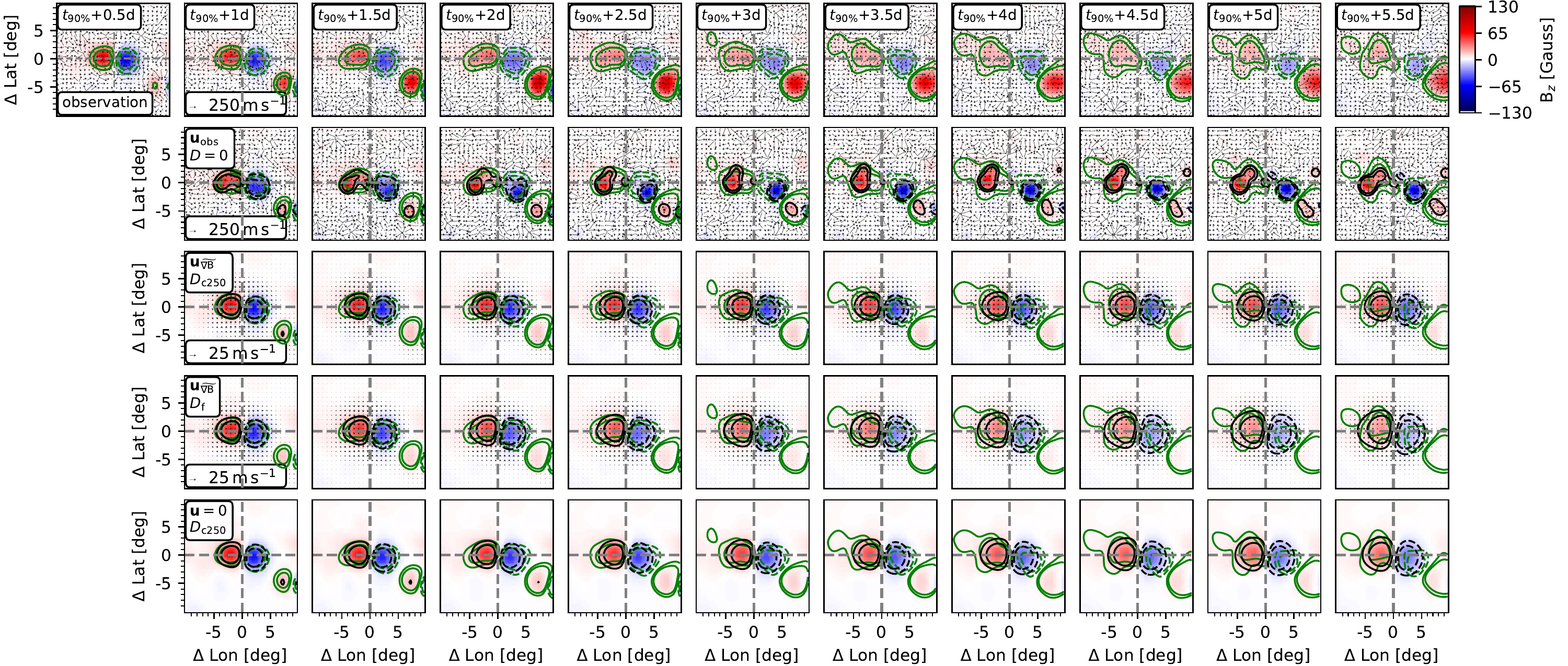}
  \caption[Simulation time steps for AR~11137]{Evolution of the magnetic field of AR~11137 in the observations (top row) and four different simulations (second row from top: no flows and constant diffusivity, third row from top: flows according to the inflow parametrization based on the gradient of the magnetic field and flux-dependent diffusivity, fourth row from top: observed flows and constant diffusivity, bottom row: observed flows and flux-dependent diffusivity). The simulations are initialized at $t_{\rm{90\%}}+0.5$~days. The times are indicated in the upper left corner of the top panels. The diffusion in the cases of constant diffusivity is \SI{250}{\kilo\meter\squared \per \second}. Reference arrows are given in the lower left corners of the second column. Red (blue) indicates positive (negative) radial magnetic field. All maps have the same saturation, at $\pm$ the rounded maximum absolute field strength in the central \SI{10}{\degree} from all simulation time steps. The green (black) contours indicate levels of half and quarter of the minimum and maximum magnetic field in the central \SI{10}{\degree} of the observation (of each simulation), for each time step individually.}
  \label{plot_examplesim_11137}
\end{figure}

\end{landscape}

Fig.~\ref{plot_examplesim_11137} shows all time steps of the observations and a few simulations, for the AR~11137. The simulations are initialized at $t_{\rm{90\%}}+0.5$~days.

\section{Average simulation}

\begin{figure*}
\centering
\includegraphics[width=\hsize]{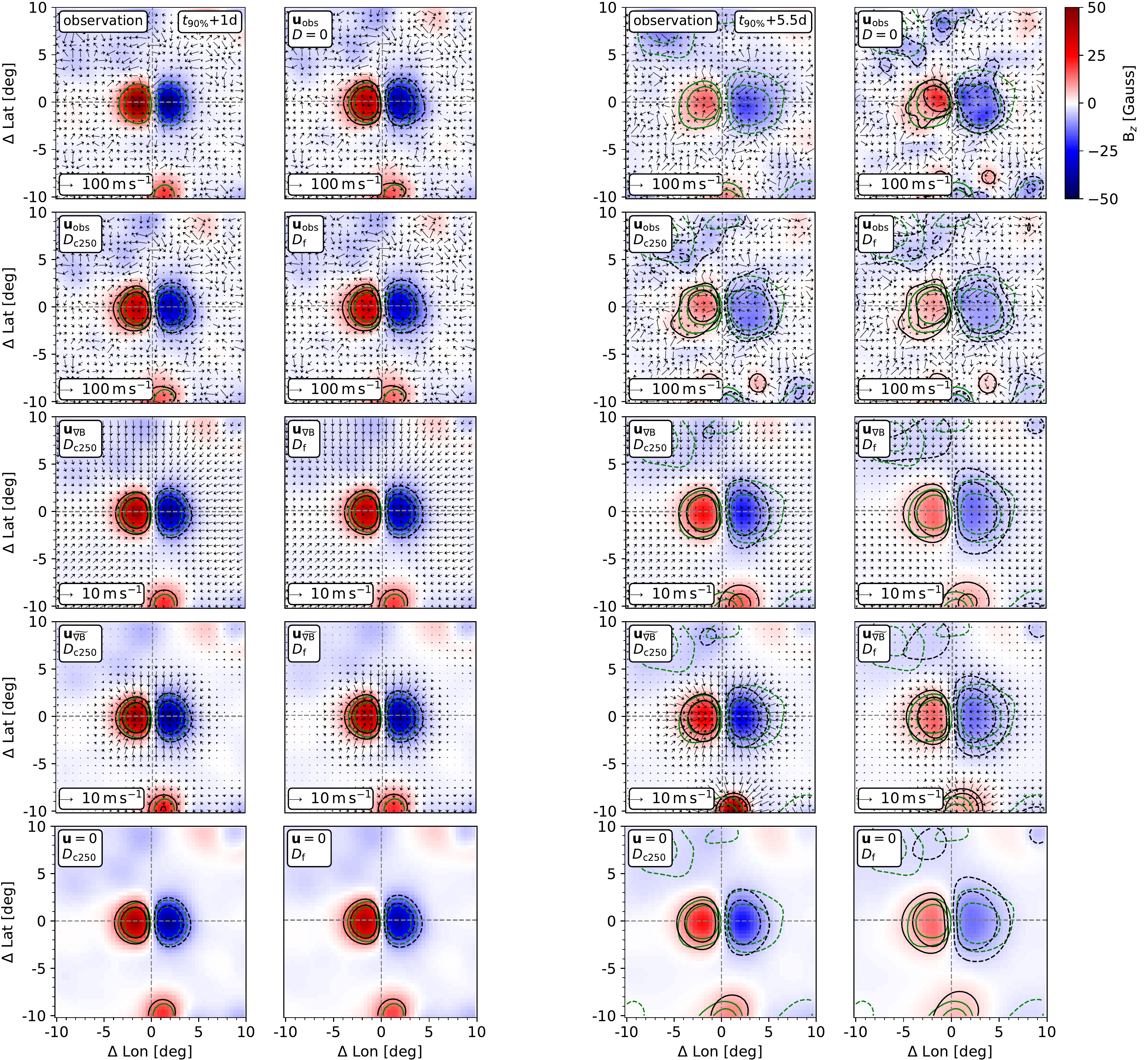}
  \caption[Average simulation time steps from ARs in the sample]{Two example time steps of the observations and the simulations, averaged over the individual ARs. The two columns on the left show the first time step after simulation start, that is, 0.5~days after $t_{\rm start} = t_{\rm{90\%}}+0.5$~days. The two columns on the right show the time step at the end of the simulations, at $t_{\rm{90\%}}+5.5$~days. The times are indicated in the upper right corner of the top left panels. At each of the two time steps, the top row shows the observed magnetic field and flows (left) and the simulation using observed flows and no additional diffusion (right). For all other rows, the left (right) panels show simulations with constant (flux-dependent) diffusivity $D_{\rm{c}}$ ($D_{\rm{f}}$). From second row to bottom: observed flows, flows according to the parameterized inflow model, flows according to the modified parameterized inflow model, and no flows. The diffusion in the cases of constant diffusivity is \SI{250}{\kilo\meter\squared \per \second}. The arrows indicate the observed and the parameterized flows, for the respective observations and simulations. Reference arrows are given in the lower left corners of each panel. Red (blue) indicates positive (negative) radial magnetic field. All maps have the same saturation, at $\pm$ the rounded maximum absolute field strength in the central \SI{10}{\degree} from all simulation time steps. The green (black) contours indicate levels of half and quarter of the minimum and maximum magnetic field in the central \SI{10}{\degree} of the observation (of each simulation), for each time step individually.}
  \label{plot_examplesim_vidframes_average}
\end{figure*}

Fig.~\ref{plot_examplesim_vidframes_average} shows two time steps of the average over the 17~ARs, for each simulation.

\section{Changing the diffusivity}
\label{appendix_diffusivities}

\begin{figure*}
\centering
\includegraphics[width=\hsize]{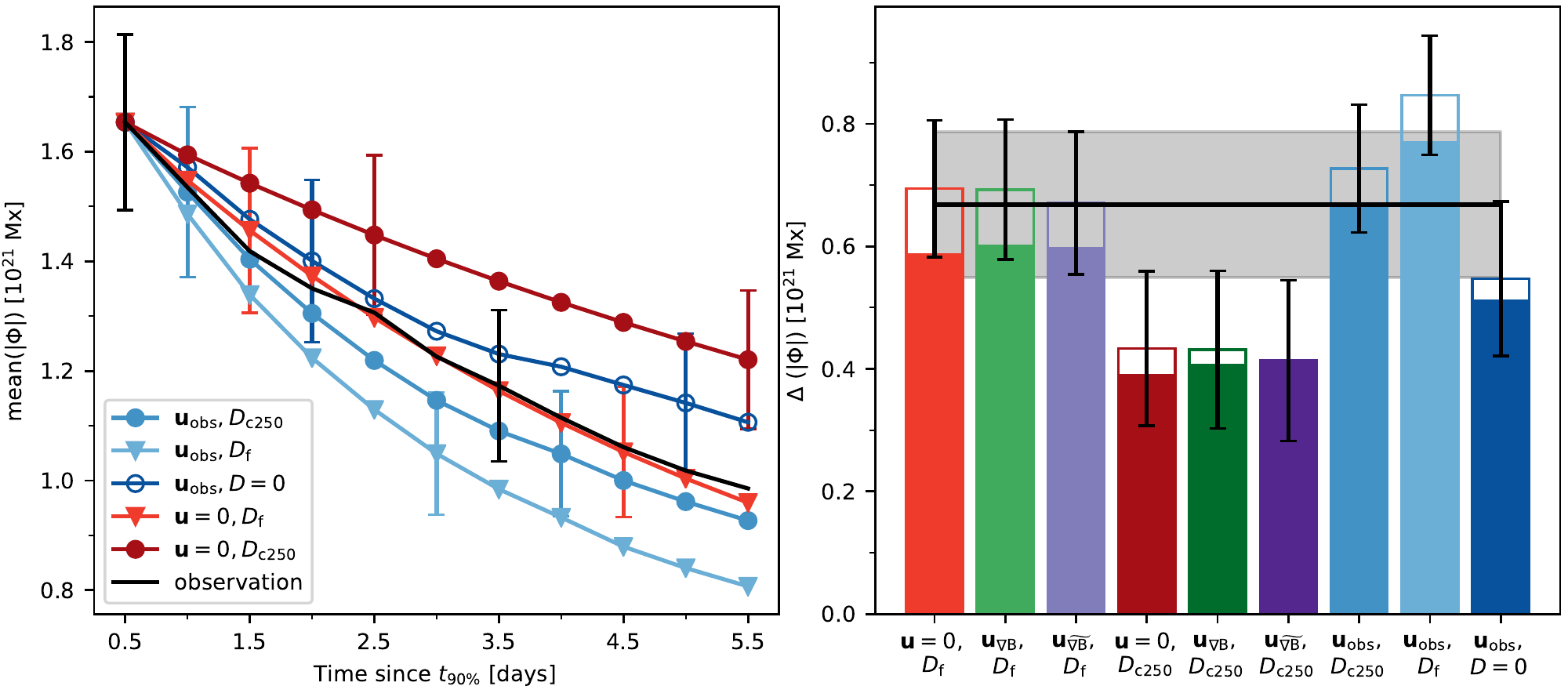}
  \caption[Evolution of AR flux in observations and simulations, at D=\SI{250}{\kilo\meter\squared\per\second}]{Left: Evolution of the total unsigned flux over the central disk with radius \SI{5}{\degree}, averaged over the sample of active regions, for the observations and some of the simulations. The error bars indicate the standard error over the sample. Only every sixth error bar is plotted, for readability. The data point at $t_{\rm{90\%}}+0.5$~days is the initial condition of the simulations. The diffusion in the simulation with constant diffusivity and no flows is $\SI{250}{\kilo \meter \squared \per \second}$. Right: The amount of flux loss between the last time step of the simulations, at $t_{\rm{90\%}}+5.5$~days, and the time when the simulations are initialized, at $t_{\rm{90\%}}+0.5$~days. The black line indicates the flux loss in the observations over the same period, the gray shaded area indicates the standard error. The solid and contoured bars indicate flux loss due to cancellation and advection, respectively.}
      \label{plot_Uflux_t90p1_sm2_vs_time_total_balance_250}
\end{figure*}

\begin{figure*}
\centering
\includegraphics[width=\hsize]{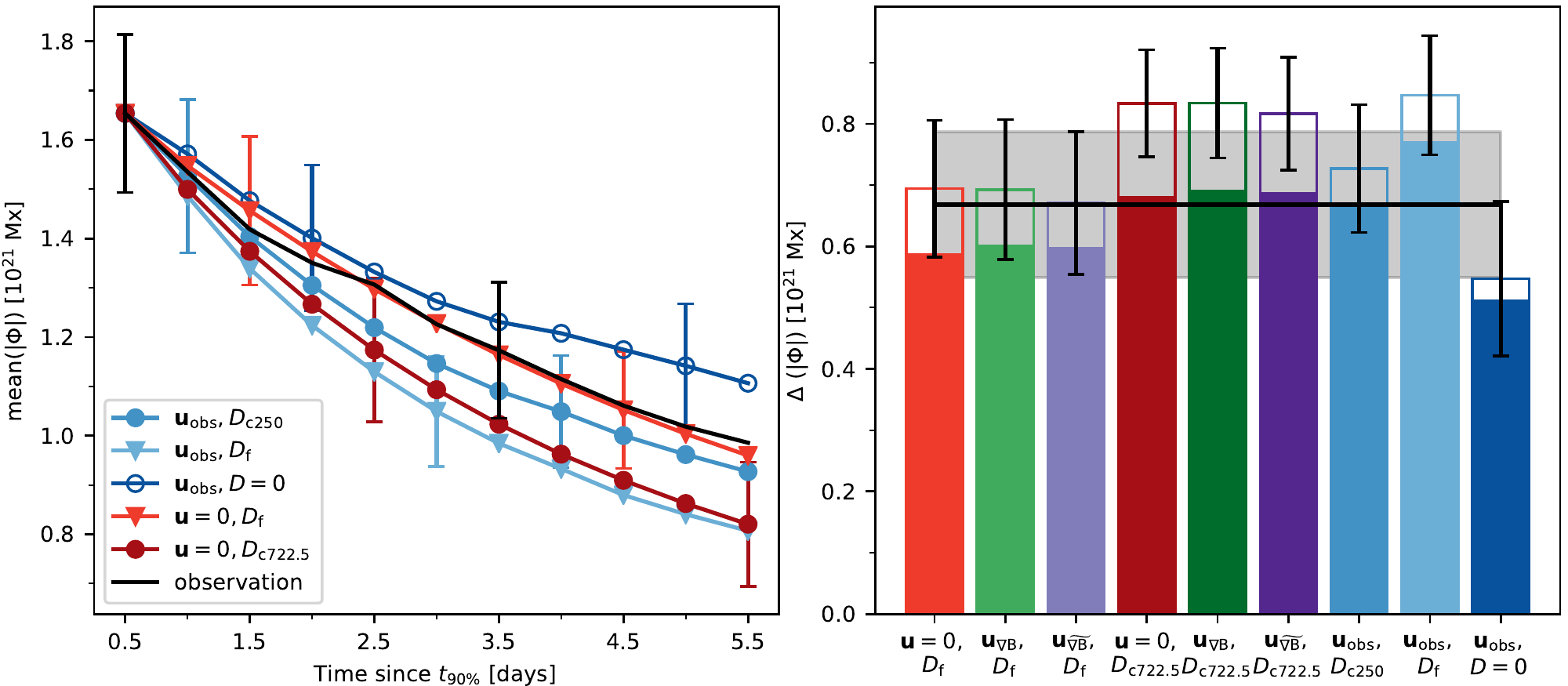}
  \caption[Evolution of AR flux in observations and simulations, at D=\SI{722.5}{\kilo\meter\squared\per\second}]{Same as Fig.~\ref{plot_Uflux_t90p1_sm2_vs_time_total_balance_250}, with a diffusion in the cases of constant diffusivity for the parameterized flow models at $\SI{722.5}{\kilo \meter \squared \per \second}$.}
      \label{plot_Uflux_t90p1_sm2_vs_time_total_balance_722p5}
\end{figure*}

Figs.~\ref{plot_Uflux_t90p1_sm2_vs_time_total_balance_250} and \ref{plot_Uflux_t90p1_sm2_vs_time_total_balance_722p5} show the total unsigned flux over the central disk with radius \SI{5}{\degree} for constant diffusivities of 250 and \SI{722.5}{\kilo \meter \squared \per \second}, respectively. The case $D_{\rm{c250}}$ corresponds to the diffusivity which serves as reference for the flux-dependent diffusivity $D_{\rm{f}}$. The case $D_{\rm{c722.5}}$ corresponds to the flux-dependent model $D_{\rm{f}}$ in the case that all corks experience no surrounding magnetic flux.

\section{Analytical solution for constant diffusion and no flows}
\label{appendix_analyticalsolultionDc}
We derive an analytical description of the evolution of the flux loss. For this, we consider the simplest case, with no flow field acting on the magnetic field ($\vec{u}=0$), and with constant diffusivity ($\mathbf{D_{\rm{c}}}$). Because diffusion acts independently per dimension, we can use a 1D setup. For the magnetic field, we consider two Gaussian distributions of opposite sign, centered at positions $\pm \Delta x$ from the origin:
\begin{equation}
    B_{\pm}(x,\sigma) = \frac{1}{\sqrt{2\pi \sigma^2}} \mathrm{exp} \Bigg\{ - \frac{(x \pm \Delta x)^2}{2 \sigma^2} \Bigg\}. \label{Gaussian1D}
\end{equation}
For small $\Delta x$, the difference between $B_{+}$ and $B_{-}$ can be written as 
\begin{equation}
    B_{+} - B_{-} = \frac{\Delta x}{\sqrt{2\pi \sigma^2}} \frac{\partial}{\partial x} \left( \mathrm{exp} \Bigg\{ - \frac{x^2}{2 \sigma^2} \Bigg\} \right). \label{Gaussian1D_difference}
\end{equation}
From this, the total unsigned flux can be calculated as 
\begin{eqnarray}
    \Phi & = & \frac{2 \Delta x}{\sqrt{2\pi \sigma^2}} \int\limits_{-\infty}^{0} \frac{\partial}{\partial x} \left( \mathrm{exp} \Bigg\{ - \frac{x^2}{2 \sigma^2} \Bigg\} \right) \mathrm{dx}\\
    & = & \frac{2 \Delta x}{\sqrt{2\pi \sigma^2}}.
\end{eqnarray}
Diffusion broadens the Gaussian distribution over time as 
\begin{equation}
    \sigma(t) = \sqrt{2 D t}.
\end{equation}
Therefore,
\begin{equation}
    \Phi (t) = \frac{2 \Delta x}{\sqrt{2\pi} \sqrt{2 D t}} = \frac{\Delta x}{\sqrt{\pi D t}}.
\end{equation}
The total flux loss over time can thus be written as
\begin{equation}
    \Delta \Phi (t) = \Phi(t_{\rm{in}}) - \Phi(t_{\rm{in}}+t) = \Phi(t_{\rm{in}}) \left( 1 - \frac{\Delta x}{\sqrt{\pi D (t_{\rm{in}} + t) }} \right), \label{eq_fluxloss_analytical}
\end{equation}
with suitable initial condition $t_{\rm{in}} = \frac{\Delta x^2}{\pi D}$.

To compare this to the simulation results, we consider an average $\Delta x = \SI{1.8}{\degree}$ (compare in Fig.~\ref{plot_examplesim_vidframes_average} the distance of the polarities from the center in the x-direction) and a time of 5~days. With a total unsigned flux of \SI{1.65e21}{\Maxwell} at the start time $t_{\rm{in}} = t_{\rm{90\%}}+0.5$~days, Eq.~\ref{eq_fluxloss_analytical} gives for the three cases of $D_{\rm{c250}}=\SI{250}{\kilo\meter\squared\per\second}$, $D_{\rm{c450}}=\SI{450}{\kilo\meter\squared\per\second}$, and $D_{\rm{c722.5}}=\SI{722.5}{\kilo\meter\squared\per\second}$ flux losses of \SI{0.39e21}{\Maxwell}, \SI{0.56e21}{\Maxwell} and \SI{0.71e21}{\Maxwell}, respectively. This is in good agreement with the flux losses due to diffusion (solid bars) in the simulations with the corresponding model of ($\vec{u}=0$), ($\mathbf{D_{\rm{c}}}$), as shown in Fig.~\ref{plot_Uflux_t90p1_sm2_vs_time_total_balance_250}, Fig.~\ref{plot_Uflux_t90p1_sm2_vs_time_total_balance}, and Fig.~\ref{plot_Uflux_t90p1_sm2_vs_time_total_balance_722p5}, respectively. The total flux loss in the observations over this time is \SI{0.66e21}{\Maxwell}.

From Eq.~\ref{eq_fluxloss_analytical}, we calculate the minimum and maximum diffusivities $D_{\rm{min}}$, $D_{\rm{max}}$ that are consistent with the observed flux loss after 5~days to a 2-$\sigma$ level. These are $D_{\rm{min}}=\SI{280}{\kilo\meter\squared\per\second}$ and $D_{\rm{max}}=\SI{1350}{\kilo\meter\squared\per\second}$, respectively. We note that this does not include flux loss due to transport away from the AR, which explains why $D_{\rm{min}}$ is larger than $D_{\rm{c250}}$.

\end{appendix}

\end{document}